\documentstyle[12pt,epsfig,amssymb]{article}
\input epsf.tex
\def\desepsf(#1 width #2){\epsfxsize=#2 \epsfbox{#1}}

\begin{document}
\def\as{\alpha_s}
\def\ee{e^+e^-}
\def\qq{q \bar{q}}
\def\lmsb{\Lambda_{\overline{\rm MS}}}
\def\to{\rightarrow}
\def\fb{~{\rm fb}}
\def\pb{~{\rm pb}}
\def\ev{\,{\rm eV}}
\def\kev{\,{\rm KeV}}
\def\mev{\,{\rm MeV}}
\def\gev{\,{\rm GeV}}
\def\GeV{\,{\rm GeV}}
\def\tev{\,{\rm TeV}}
\def\beq{\begin{equation}}
\def\eeq{\end{equation}}
\def\beqn{\begin{eqnarray}}
\def\eeqn{\end{eqnarray}}
\def\3g2ga{$ggg\gamma\gamma$}
\def\nn{\nonumber\\}

\setlength{\parskip}{0.45cm}
\setlength{\baselineskip}{0.75cm}
\begin{titlepage}
\begin{flushright}
ETH-TH/99-15 \\ 
\end{flushright}
\vspace{0.1cm}
\begin{center}
\Large{ Two photons plus jet at LHC: the NNLO \\  contribution from the $gg$ initiated process
 \footnote{Work partly supported by the EU Fourth Framework Programme `Training and Mobility of Researchers', Network `Quantum Chromodynamics and the Deep Structure of Elementary Particles', contract FMRX-CT98-0194 (DG 12 - MIHT).}}

\vspace{1.cm}
\large{D.\ de Florian and Z.\ Kunszt\\}
\vspace{0.35cm}
\normalsize
{\it Institute of Theoretical Physics, 
ETH \\ CH-8093 Z\"urich, Switzerland}

\hspace{1cm}

\large
{\bf Abstract} \\
\end{center}
\vspace{0.5cm}
\normalsize

The production
of the Standard Model Higgs boson of mass 
$\approx 100-140\gev$  at the LHC likely gives  clear signals
in the $\gamma  \gamma$ (1) and
in the $\gamma   \gamma  $ jet  (2)  channels. 
The quantitative evaluation of the
background to channel (1) is very hard since
the next-to-leading order (NLO) and next-to-next-to-leading order (NNLO) QCD 
corrections are large. 
 In particular, the contributions of the NNLO QCD subprocess $ gg\to \gamma  
 \gamma $ to inclusive $ \gamma\gamma $ production is comparable to the 
 contribution of the leading order subprocess $ q\bar{q}\to \gamma \gamma $.
The quantitative description of the background to channel (1), therefore,
requires to calculate all corrections up to the NNNLO level.
In this letter we
 present results  on the contribution of the 
 NNLO QCD subprocess $ gg\to g \gamma  \gamma $
to the production rate of channel (2).
We have found that in this case 
this NNLO contribution is less than 20\% of the Born contribution. 
Since the NNLO contributions will likely be dominated by this
subprocess  one can argue that in the case of channel (2) - contrary
to the case of channel (1) -  
 a quantitative  description of the  background can be achieved already 
at next-to-leading order accuracy.

\end{titlepage}
\newpage

The search for Standard Model Higgs boson will be a high priority
project at the LHC collider. Recent results of the LEP experiments obtained from
a fit to the precision data and by direct search
 indicate that the mass of the Higgs boson  at the 95\% confidence
level should be in the mass range of $90$ GeV 
$\leq m_H \leq 262$ 
GeV \cite{cern}.
However, the search for the Higgs boson at the LHC  is rather difficult
 in the low mass range of $100$ 
GeV 
$\leq m_H \leq 140$ 
GeV when measurable
signals exist for the two photon decay channel.
The simulations carried out by ATLAS and CMS \cite{lhc} have shown
 that the significance of the signal of Higgs
production  in the inclusive channel 
$pp\rightarrow \gamma\gamma $ is rather low.
Assuming 
integrated luminosity of $3\cdot 10^{-4}\,{\rm pb}^{-1}$ 
one obtains   $S/\sqrt{B}\approx 5\, - 10\, $, the actual value
depending on the Higgs mass and the assumptions  
 on the size of the higher order QCD corrections.

 Next-to-leading order corrections to the signal are 
known to be large \cite{spira} 
(with a $K$-factor of $\approx 1.8$), 
 whereas the full calculation of the background, including fragmentation effects, 
 at the same order is just being evaluated \cite{guillet}.
The NLO results, however,  will likely
 be not enough for having a quantitative description of
 the background.
 At LO only  the $q\bar{q}\rightarrow \gamma\gamma$
 subprocess can contribute and, for this 
initial state,  the parton-luminosity is rather low. 
It is not surprising, therefore, that the
  one-loop subprocess $gg\rightarrow \gamma\gamma$ (box diagram) \cite{dicus}, 
appearing first in NNLO accuracy 
but having high initial parton-luminosity
and  collinear singularity structure similar to  a Born one,
gives production-rates which  are numerically
very similar (both in size and shape) 
to the LO result.
This poses a serious theoretical problem,
 since  in order to achieve a significant reduction in the scale dependence one has to go
 one order further, that is to NNNLO,
 which is completely beyond the scope of the available techniques.

Recently, it has been pointed out \cite{adikss}
that the  production of the Higgs boson in association with a hard jet,
 and its subsequent decay to a photon pair, also gives measurable signal. 
 This channel  has both experimental and theoretical advantages. 
The detection of the more energetic photons in association with a hard  
jet allows to improve the efficiency and the
two photon invariant mass resolution. Furthermore, the 
existence of a jet in the final state allows for the implementation 
of different cuts which can be particularly useful to suppress the background. 
The cross-section of the signal process
 $p p\rightarrow H $ jet  has  recently been computed to  NLO 
accuracy, in the limit of $m_H\leq m_t$, finding 
 a $K$-factor similar to the more
 inclusive case $K\sim 1.6$ \cite{dgk}. 
Although all the NLO matrix elements
needed to construct the corresponding
NLO Monte Carlo program for numerical evaluation of the background are available \cite{signer}, such a program has not been implemented yet.
Furthermore, the theoretical problem of the gluon-gluon initial state pointed out for the inclusive channel may also be an issue in this case.
In the Born approximation, the background 
receives contributions from the  subprocesses
$q\bar{q} \rightarrow g\gamma\gamma$ and  $qg \rightarrow q \gamma\gamma$. The 
subprocess with gluon-gluon initial state
$gg\rightarrow g\gamma\gamma$ (box and pentagon diagrams)
contributes   again first in NNLO accuracy. But there is an   important 
difference: in this case, subprocesses with gluon-quark initial state having a high parton-luminosity contribute already at the lowest order. 
It is, therefore,  expected  
that the  NNLO $gg$ contribution will be 
far less significant than in the inclusive case.
 If this expectation turns out to be correct,
 then one may argue that 
 a quantitative  description of the  background can be achieved already 
at next-to-leading order accuracy. 
  The evaluation of this  $gg$ initiated
NNLO correction for the production
 of two photons plus one jet at the LHC is the main purpose of this letter.

The one loop   $3g2\gamma$ amplitudes
have not been directly computed. But fortunately,   
the  one loop results 
of Bern, Dixon and Kosower  \cite{dixon} on the $5g$ amplitudes
 have been documented   in terms of partial amplitudes
with explicit  color decomposition;  therefore one can obtain  the   $3g2\gamma$ amplitudes by changing
color factors, normalizations and doing appropriate permutations \footnote{We  
  thank L. Dixon for his assistance concerning these modifications.}.

The $n_f$ part of the one loop $5g$ amplitude can be decomposed in the following way
\beqn
\label{5g}
{\mathcal{A}}_{5\,\, [n_f]}^{\rm 1-loop}= g^5 \sum_{\sigma\in S_5/Z_5} {\rm Tr}\left[ T^{a_{\sigma(1)}} T^{a_{\sigma(2)}} T^{a_{\sigma(3)}} T^{a_{\sigma(4)}} T^{a_{\sigma(5)}} \right] A_{5;1}^{[1/2]}(\sigma(1),\sigma(2),\sigma(3),\sigma(4),\sigma(5))
\eeqn
where $S_5/Z_5$ is the set of the 24 non-cyclic permutations of five objects and $A_{5;1}^{[1/2]}$ is the partial amplitude corresponding to the case of a fermion circulating in the loop. The $T^a$ are the SU(3) generators in the fundamental representation, normalized so that ${\rm Tr}(T^a T^b)=\delta^{ab}$.

The one loop $3g2\gamma$ amplitude (with the photons labeled by $4$ and $5$)
 has the same  color decomposition and it can be obtained \cite{signer} by replacing in eq.(\ref{5g}) the corresponding
 SU(3) generators of the gluons by the U(1) values (for each photon) and 
 the QCD coupling $g$ by the electric charge of the quark circulating within the loop, as 
\beqn
T^{a_4} \rightarrow \sqrt{2} \nn
T^{a_5} \rightarrow \sqrt{2} \nn
g^5 \rightarrow g^3 e_Q^2 .
\eeqn
In  the color decomposition for the 3 remaining gluons, after these replacements, only two color structures appear, i.e
\beqn
{\mathcal{A}}^{\rm 1-loop}_{3g2\gamma}(1,2,3;4,5)&=& g^3 \left( e_Q \sqrt{2}\right)^2 \\
 &&\left\{ {\rm Tr}\left[ T^{a_1}T^{a_2}T^{a_3}\right] A_5^{2\gamma} (1,2,3;4,5)
 + {\rm Tr}\left[ T^{a_3}T^{a_2}T^{a_1}\right] A_5^{2\gamma} (3,2,1;4,5) \right\} \nonumber
\eeqn
where $A_5^{2\gamma} (1,2,3;4,5)$ is obtained by collecting the 12 permutations of $A_{5;1}^{[1/2]}$  with the right cyclic ordering of $(1,2,3)$,
\beqn
A_5^{2\gamma} (1,2,3;4,5) &=& A_{5;1}^{[1/2]}(1,2,3,4,5) +  A_{5;1}^{[1/2]}(1,2,3,5,4) +  A_{5;1}^{[1/2]}(1,2,4,3,5)\nn 
&+&  A_{5;1}^{[1/2]}(1,2,5,3,4)  +  A_{5;1}^{[1/2]}(1,2,4,5,3) +  A_{5;1}^{[1/2]}(1,2,5,4,3) \nn
&+&  A_{5;1}^{[1/2]}(1,4,2,3,5) +  A_{5;1}^{[1/2]}(1,4,2,5,3) 
+ A_{5;1}^{[1/2]}(1,4,5,2,3)\nn
 &+&  A_{5;1}^{[1/2]}(1,5,2,3,4) +  A_{5;1}^{[1/2]}(1,5,2,4,3) +  A_{5;1}^{[1/2]}(1,5,4,2,3). 
\eeqn
Furthermore,  by using Furry's theorem one obtains
\beqn
 A_5^{2\gamma} (1,2,3;4,5) = - A_5^{2\gamma} (3,2,1;4,5).
\eeqn

The partial amplitudes $A_{5;1}^{[1/2]}$ for the  four independent helicity structures $(+++++),(-++++),(--+++),(-+-++)$ are given in ref. \cite{dixon}. The remaining ones are obtained by cyclic permutations or complex conjugation. The same method can be used to re-compute the $2g2\gamma$ amplitudes from the $4g$ ones evaluated in ref. \cite{zoltan}.
We have verified numerically soft and collinear limits 
for  the $A_5^{2\gamma}$ helicity amplitudes as well as for their spin averaged squared values. 

The cross-section values presented below have been obtained using the LO MRST \cite{pdf} parton distributions and with  the factorization and renormalization scales given by
\beqn
  Q_0^2= M_{\gamma\gamma}^2+ p_{T(jet)}^2,
\eeqn
 where $M_{\gamma\gamma}$ corresponds to the invariant mass of the photon pair. The jet and the photons are required to have $p_T>40$ GeV and rapidity within $|\eta|<2.5$, i.e., corresponding to the  {\bf C1} cuts introduced in the analysis of ref. \cite{adikss}. Furthermore, since photons should be isolated, we apply a cut $\Delta R= \sqrt{\Delta \phi^2+\Delta\eta^2}>0.3$ for each pair of particles, avoiding in this way any possible $q\gamma$ final state collinear singularity.
In this work, we are only interested in the direct production of two photons, so we neglect any contribution coming from the single and double fragmentation processes. Since the $\gamma\gamma$ measurement is intended for a detection of a Higgs boson lighter than the top quark, we assume only 5 flavors and consider all of them massless. This has been 
shown \cite{dicus} to be a very good approximation for the box contribution, where the massive expressions are also known \cite{constantini}.

In order to make comparison with the case of the  inclusive channel $pp\rightarrow \gamma\gamma $ we have also evaluated the corresponding $q\bar{q}\rightarrow \gamma\gamma$ and $gg\rightarrow \gamma\gamma$ contributions with similar cuts for the photons  ($p_T>40$ GeV, $|\eta|<2.5$) and $Q_0^2= M_{\gamma\gamma}^2$ as default scale.

\begin{figure}[htb]
\centerline{  \desepsf(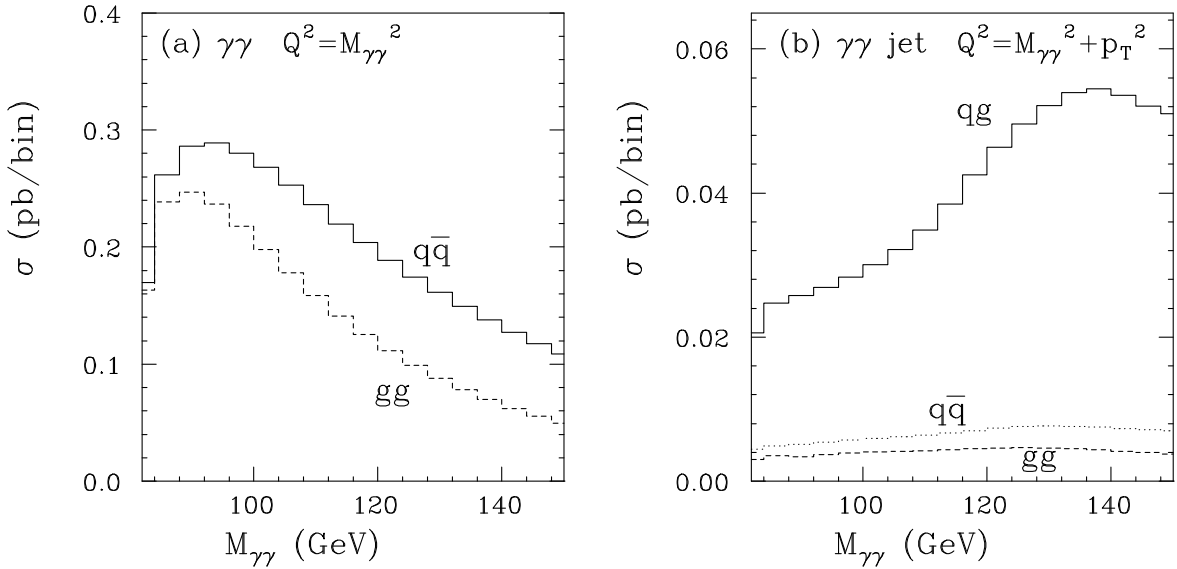 width 18cm) ~~~~~~~~~} 
\caption{Contributions of the different partonic initial states to the 
cross-section for processes (a) $pp\rightarrow \gamma\gamma $ and (b) 
$pp\rightarrow \gamma\gamma$ jet. } 
\end{figure}
In Fig. 1(a) we show the results for the inclusive channel, from where one can see that the NNLO  ($gg$) contribution is rather large and actually it is comparable to the lowest order ($q\bar{q}$) one, as stated before.
In Fig. 1(b) we show the lowest order ($qg$ and $q\bar{q}$) and  NNLO ($gg$) contributions for the case of $\gamma\gamma $ jet production. It is clear from the plot that, as in the inclusive case, the NNLO $gg$ initiated contribution is similar both in magnitude and shape to the  $q\bar{q}$ initiated one. But since  the Born cross section is  dominated by the $qg$ subprocess, the $gg$ NNLO term becomes less significant 
than for the inclusive channel.
 Of course, since all these contributions for each partonic initial state are  `Born-like' ones, the results presented in Fig. 1 are strongly scale dependent. In order to study the effect of using  different scales, in Fig. 2 we show the  scale dependence of the ratio between the NNLO $gg$ initiated and the Born contributions $R=\sigma^{gg}/\sigma^{LO}$, for both the inclusive and the jet channel and for 3 different values of the invariant mass of the photon pair $M_{\gamma\gamma}=100, 120, 140$ GeV. We vary the default scales by introducing a parameter $\mu$ such that $Q=\mu\, Q_0$, where $Q_0$  corresponds to the default scale defined above for each process. As it can be observed from Fig. 2,  even though the scale dependence for the ratios are quite similar, in the jet channel the $gg$ contribution remains always smaller than 20\% of the Born one, whereas for the inclusive channel it can even be larger than the LO term at the lower scales.

\begin{figure}[htb]
\centerline{  \desepsf(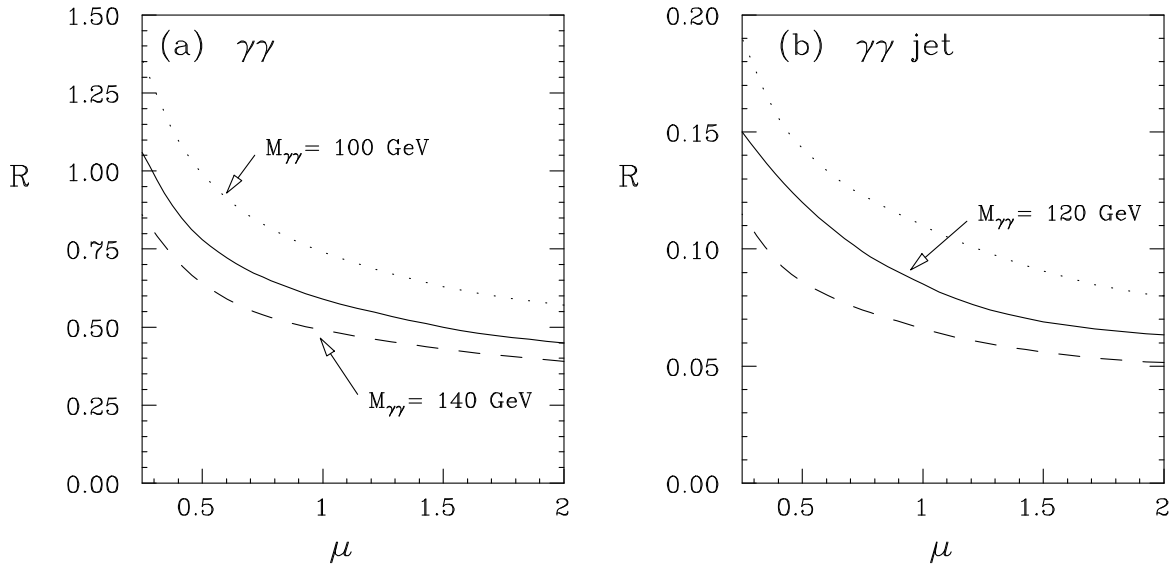 width 18cm) ~~~~~~~~~} 
\caption{Ratio of the NNLO $gg$ initiated contribution over the LO one for (a) $pp\rightarrow \gamma\gamma $ and (b) $pp\rightarrow \gamma\gamma$ jet. } 
\end{figure}

In conclusion,  we have 
 presented results  on the contribution of the 
 NNLO QCD subprocess $3g2\gamma$
to the production-rate of two photons plus one jet at the LHC.
We have found that  
this NNLO contribution is less than 20\% of the Born contribution. 
Since the NNLO contributions will likely be dominated by this
subprocess,  one can argue that in the case of associated production
of a Higgs boson with one jet  
 - contrary
to the case of the completely inclusive production  - one can achieve  
 a quantitative  description of the  background  already 
in next-to-leading order accuracy. 
This provides us another argument for favoring the $\gamma\gamma$ jet signal in searching for the Standard Model Higgs boson in the mass range of $m_H \approx 100 - 140$ GeV.

We are grateful to L. Dixon, M. Grazzini and A. Signer for helpful discussions.


\end{document}